\documentclass[lettersize,journal]{IEEEtran}
\usepackage{amsmath,amsfonts}
\usepackage[ruled, lined, longend, linesnumbered]{algorithm2e}
\usepackage{array}
\usepackage[caption=false,font=normalsize,labelfont=sf,textfont=sf]{subfig}
\usepackage{textcomp}
\usepackage{stfloats}
\usepackage{url}
\usepackage{verbatim}
\usepackage{graphicx}
\usepackage{cite}
\usepackage{multirow}
\begin{document}

\title{A Novel Local and Hyper-Local Multicast Services Transmission Scheme for Beyond 5G Networks}

\author{Sweta Singh and K. Giridhar
\thanks{The authors are with the Department of Electrical Engineering, Indian Institute of Technology, Madras, Chennai, 600036, India (email:singhsweta@telwise-research.com, k.giridhar@telwise-research.com)}
}

\markboth{Journal of \LaTeX\ Class Files,~Vol.~14, No.~8, August~2021}%
{Shell \MakeLowercase{\textit{et al.}}: A Sample Article Using IEEEtran.cls for IEEE Journals}

\IEEEpubid{0000--0000/00\$00.00~\copyright~2021 IEEE}

\maketitle

\begin{abstract}
The efficiency of the broadcast network is impacted by the different types of services that may be transmitted over it. Global services serve users across the entire network, while local services cater to specific regions, and hyper-local services have even narrower coverage. Multimedia Broadcast over a Single-Frequency Network (MBSFN) is typically used for global service transmission while existing literature extensively discusses schemes for transmitting local or hyper-local services with or without Single Frequency Network (SFN) gain. However, these schemes fall short when network-wide requests for only local and hyper-local services are made, leading operators to scale down to either Single Cell-Point to Multipoint (SCPtM) or Multi-Frequency Network (MFN). SCPtM is highly susceptible to interference, and MFN requires substantial amounts of valuable spectrum. They both employ the Least Channel Gain (LCG) strategy for transmitting hyper-local services without SFN gain. 

Our proposed Local and Hyper-Local Services (LHS) transmission scheme utilizes the knowledge of user distribution and their corresponding radio link channel quality to schedule single or multi-resolution, local or hyper-local services within a three-cell cluster and aims to enhance spectral efficiency and maximize system throughput. It leverages Scalable Video Coding (SVC) in conjunction with Hierarchical Modulation (HM) for transmitting multi-resolution multimedia content to address the problem of heterogeneity amongst the multicast group users. The proposed scheme also employs macro-diversity combining with optimal HM parameters for each gNB catering to a local service area in order to minimize the service outage. System-level simulation results testify to the better performance achieved by the proposed LHS transmission scheme with respect to SCPtM.
\end{abstract}


\section{Introduction}
\IEEEPARstart{M}{ulticasting}/Broadcasting or Point to Multipoint (PtM) helps minimize the network load, compared to point-to-point data transfers, when a service is requested simultaneously by many users in a particular region. The demand for multicast applications in cellular systems is on a rapid rise, with global average mobile data usage per smartphone projected to increase from 21 GB in 2023 to 56 GB in 2029. The share of 5G in mobile data traffic is anticipated to reach 76 percent by 2029, emphasizing the significant role of this technology \cite{Ericsson}. Video content carried by smartphones and other devices will dominate mobile broadband traffic, and the International Telecommunication Union (ITU) predicts that it will be 4.2 times greater than non-video in 2025 and 6 times greater in 2030 \cite{series2015guidelines}. 

The surge in enhanced video services, from downloads to streaming and a diverse array of entertainment, interactive, and real-time applications, underscores the pathway to future 5G human-oriented multicasting. Such applications demand high data rates, low jitter, and ubiquitous connectivity, especially in mobile scenarios. Another category of applications focuses on transmitting user-customized information, including news, advertising, location-based services like augmented reality multicast applications tailored for commercial and tourist services, and public safety transmissions during emergencies. These applications demand low-latency data transmission, high reliability, and extended coverage \cite{7853753}.

Multicast traffic goes beyond end-user device groups, including machine-type communications (MTC) for the Internet of Things (IoT). Therefore, there is a necessity to investigate efficient methodologies that can support the delivery of real-time and on-demand video content.

The Third-Generation Partnership Project (3GPP) initially standardized Multimedia Broadcast/Multicast Service (MBMS) and later evolved it to eMBMS. Subsequently, the development of 5G New Radio (NR) by 3GPP introduced advanced capabilities to handle numerous service requests and accommodate various Multicast/Broadcast applications. These applications include eMBMS, location/position-based, critical communication, and Vehicle-to-Everything (V2X) services. To efficiently support these services, the International Telecommunication Union (ITU) defined three usage scenarios: enhanced Mobile Broadband (eMBB), massive Machine Type Communications (mMTC), and Ultra-Reliable and Low Latency Communications (URLLC). \cite{8886714} Our current work focuses explicitly on the eMBB scenario, utilizing periodic user Channel Quality Indicator (CQI) feedback to determine the Modulation and Coding Scheme (MCS) level for delivering multicast traffic via eMBMS.
\subsection{Service Types}
Certain services, like nationwide TV programs, attract a large audience and thus are suitable for broadcasting across the entire network, making them global services. Conversely, some services are better suited for efficient broadcasting only within specific sub-regions of the network, such as a city. For these services, broadcasting within the sub-region is practical, but point-to-point transmission might be more suitable outside this area. These services are termed local services, and the \\

region they cover is known as the local service area \cite{4380261}. 

\begin{figure}[!t]
\centering
\includegraphics[width=2.2in]{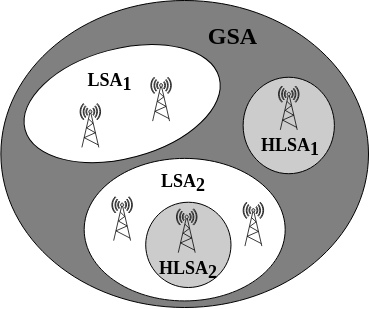}
\caption{Different types of Service Areas: Global Service Area (GSA), Local Service Area (LSA) and Hyper-Local Service Area (HLSA).}
\label{label:serviceAreas}
\end{figure}

In certain instances, the service area is even smaller than the local service area, and we will refer to such services as hyper-local services, with their corresponding transmission area termed as a hyper-local service area. While traditional MBMS transmission networks commonly categorize both local and hyper-local services as local services, we retain this distinct sub-classification for clarity and precision. The fact that each of these services has a different target region as shown in Figure [\ref{label:serviceAreas}], has to be considered when designing a digital broadcast network to deliver all these services efficiently.

\subsection{Current Schemes for MBMS Transmission} 
 Multimedia Broadcast/Multicast Service Single Frequency Network (MBSFN), as specified in 3GPP Release 9, operates as a reuse-1 system utilizing a simulcast transmission technique. It involves the simultaneous transmission of identical waveforms on the same frequency band from multiple cells within a large, statically configured MBSFN area. The User Equipment (UE) perceives the MBSFN transmission from multiple cells as a single transmission, and this diversity gain allows it to operate at a high Signal-to-Noise Ratio(SNR) \cite{8531725}. It is well-suited for global services due to its more efficient utilization of available bandwidth, simplified radio-planning process, and fault-tolerant coverage performance. Utilizing a Single Frequency Network (SFN) for disseminating local services such as local news, location-based applications, and targeted advertisements would necessitate transmission by all transmitters, including regions where they are not needed, resulting in significant capacity wastage. 

The literature suggests various mechanisms for local service insertion (LSI), which can be broadly categorized into two types: LSI while preserving the traditional Single Frequency Network (SFN) gain and LSI without SFN gain. The schemes with SFN gain include orthogonal time-division-multiplexing (TDM) or TS-LSI, orthogonal frequency-division-multiplexing (FDM) or OLSI \cite{6812187}, multiple-input-multiple-output (MIMO) technologies \cite{7185362}, Layered-Division-Multiplexing (LDM) \cite{6222350} while those offering LSI without SFN gain include Hierarchical Local Service Insertion (HLSI) \cite{6823709}, Bit Division Multiplexing (BDM) \cite{6509443}, Single Cell Point to Multipoint (SCPtM) \cite{scptm} and Multi Frequency Networks (MFNs) \cite{4380261}. 

O-LSI allocates a set of OFDM sub-carriers for local service transmission, with each LSA transmitter using a subset of these sub-carriers.

TS-LSI allocates a subset of time slices for local content, offering high simplicity. Reserved time slots for local services prevent interference with global services, preserving SFN gain. However, the coverage area for local services is smaller than in an MFN approach due to interference between different local service areas. Despite its advantages, OLSI and TS-LSI exhibit lower capacity and spectral efficiency compared to ideal superposition coding \cite{4380261}.

Employing MIMO introduces receiver complexity and demands favorable Signal-to-Noise Ratio (SNR) conditions for achieving optimal local service capacity. 

Layered-Division-Multiplexing (LDM), also known as Cloud Transmission, utilizes a multi-layer signal structure to transmit multiple broadcasting services simultaneously over the same frequency band, with layers allocated different transmission powers and signal configurations \cite{7378924}. Compared to traditional TDM/FDM systems, LDM significantly increases capacity within a single radio frequency (RF) channel while ensuring no coverage gaps among adjacent local service areas. No directional receiving antenna is needed, as receivers simply tune into the stronger local received signal \cite{6746098}.

HLSI employs Hierarchical Modulation (HM), generating QAM symbols from two bit streams with varying robustness levels. Global content is transmitted in the High Priority (HP) stream, and local content is inserted into the Low Priority (LP) stream. While hierarchical modulation offers low-complexity receivers, it results in limited throughput for local services and suffers from constrained flexibility in channel resource allocation.

Bit Division Multiplexing (BDM) leverages the inherent Unequal Error Protection (UEP) characteristic in high-order constellation mapping. It allocates bit resources across multiple symbols for multi-service transmission, dividing channel resources at the bit level for global and local services. This approach offers higher flexibility than the conventional Hierarchical Modulation (HM) strategy, albeit with increased transmitter and receiver complexity \cite{6509443}.

The 3GPP proposed an alternative transmission mode to MBSFN to support hyper-local services while avoiding significant capacity waste. SC-PtM, popularly known as the Universal Frequency Reuse (UFR) system, operates on a reuse-1 principle, randomly assigning Physical Resource Blocks (PRBs) to each Multicast Group requesting multimedia content. However, this random allocation poses challenges, especially at the cell edge, where users may experience severe Inter-Cell Interference (ICI), resulting in lower throughput. The only difference between SCPtM and MFN is that in the latter, each transmitter serves a network cell with a unique transmission frequency, reducing interference from nearby cells. 

However, the MBMS techniques described above are spectrally inefficient as they utilize the Least Channel Gain (LCG) strategy. The following section elaborates on the challenge of heterogeneity and the current techniques in the literature employed to address it.

\subsection{Problem of Heterogeneity}
Multiple multicast groups exist in a cell, and each multicast group may contain users located at diverse locations within a cell. Thus, each user may experience a different signal-to-noise ratio (SNR), which determines the maximum rate at which this user can receive data reliably. \cite{Won}  

The primary difficulty in multicast scheduling arises from this mismatch in data rates among users within a multicast group, commonly known as the problem of heterogeneity. As all users in a multicast group must adhere to the same transmission rate set by the base station at each time slot, finding a suitable rate for the entire group is challenging. Transmitting at the highest rate requested by users may lead to some users missing the transmission, while using the lowest rate may result in users with better channel conditions receiving sub-optimal rates. Thus, to meet Quality of Service (QoS) requirements, the Least Channel Gain (LCG) strategy, while being spectrally inefficient, adjusts the Modulation and Coding Scheme (MCS) based on the capabilities of the weakest terminal in the multicast group. \cite{Won, Radhakrishnan2012, 8531725}

The DA-TV broadcast technique, as described in \cite{9027462}, adapts the MCS parameter and transmits the best available video quality level to different sub-groups of users within a multicast group to boost their Quality of Experience. The authors in \cite{9069443} partition a given MBSFN area into different sub-regions to deliver the same service with different quality levels determined by the bottleneck users.

Scalable Video Coding (SVC), as explained thoroughly in the next section, provides an attractive option for sending the same multimedia content in multiple layers as base and enhancement layers, improving the spectral efficiency of the system \cite{6148193}. Receiving the base layer allows cell-edge or Least Channel Gain (LCG) users to access Standard Definition (SD) video, and incorporating enhancement layers enhances video quality for High Channel Gain (HCG) users, delivering Higher Definition (HD) videos to them. \cite{Radhakrishnan2012}

Subgrouping in conjunction with SVC has been popularly studied by many. In \cite{8886714}, the authors propose a Dynamic MBSFN Area Formation (DMAF) algorithm that MBSFN Areas by leveraging the multicast subgrouping paradigm for SVC traffic delivery.  

\subsection{Motivation}
SFN is ideal when the whole cluster has no Local Services request made.
However, if Local Service requests are being made such that the number of Local Services is significantly lower than the Global Services being offered, then it is sensible to use OLSI. OLSI decreases the available capacity of these Global Services with every new Local Service being added since they both share all available OFDM data sub-carriers. While the Global Services still experience SFN gain in the overlapping zones between adjacent Local Service Areas, Local Services reception is interference-free. Also, with no SVC deployed for Global and Local Service transmission, the problem of heterogeneity persists. 

HLSI becomes suitable when the number of local services is considerably lower than that of global services or, at most, equal. Local Services coverage is confined to the proximity of the gNB, introducing Inter-Layer Interference (ILI) to Global Services transmitted earlier. The High-Priority (HP) stream can be 16 QAM or QPSK, while the Low-Priority (LP) stream is consistently set to QPSK, ensuring the multiplexed local services are Standard Definition (SD). To mitigate Inter-Cell Interference (ICI) from adjacent Local Service Areas (LSAs), similar to SCPtM, the authors \cite{olsi} propose using Iterative Sliced Decoding (ISD) and deploying time-sharing between distant LSAs.

Single-Frequency Network (SFN) forms the baseline network structure for both HLSI and OLSI, and the local services inserted can either be hyper-local or local, depending on the service area. While local services experience diversity gain based on the number of cells in the Local Service Area (LSA), hyper-local services (HLSs) do not. The baseline network structure for both HLSI and OLSI is Single-Frequency Network (SFN).

However, in situations where no Global Services(GSs) need to be transmitted, network service providers opt for scaling down to either Single Cell-Point to Multipoint (SCPtM) or Multi-Frequency Network(MFN). As previously discussed, SCPtM is highly susceptible to interference, and MFN requires substantial amounts of valuable spectrum. Both networks utilize the Least Channel Gain (LCG) strategy for transmitting hyper-local services and lack the Single-Frequency Network (SFN) gain.

This paper introduces a novel Local and Hyper-Local Services (LHS) transmission scheme tailored for a statically configured cellular network aimed at mitigating the limitations of SCPtM. Our key contributions include: (i) Optimizing radio resource allocation for the requested multimedia contents to efficiently enhance spectrum utilization. (ii) Leveraging Scalable Video Coding (SVC) in conjunction with Hierarchical Modulation (HM) for efficiently transmitting multi-resolution multimedia content. (iii) Selecting optimal HM parameters for each gNB catering to a local service area.

The proposed LHS transmission scheme aims to optimize system throughput by transmitting the most requested multimedia content within a three-cell cluster as local services. The utilization of macro-diversity combining in conjunction with Scalable Video Coding (SVC), helps minimize service outage for these local services. The spectrum efficiency of our proposed scheme is further enhanced by transmitting either multi-resolution or single-resolution hyper-local multimedia services based on the knowledge of user distribution. To the best of our knowledge, this problem of structurally transmitting hyper-local services alongside local services has not been rigorously investigated in the literature.

The paper is structured as follows: Section II delves into the intricacies of the proposed LHS transmission scheme and the underlying scheduling and $\alpha$ assigning algorithms. Section III presents and discusses the simulation results. The concluding remarks, along with insights for future work, are summarized in Section IV.

\section{Proposed LHS Scheme}
The proposed LHS transmission scheme categorizes multimedia content requests across the network into two main groups based on the reception area and the content requests being made: Local services and Hyper-Local services. Video Layering, followed by hierarchical modulation of the multimedia content to be transmitted, irrespective of its category, is utilized to transmit multi-resolution multimedia content and address the problem of heterogeneity among the multicast group users thereby enhancing users’ QoE and fairness\cite{1054727, 210540}. 

Local service transmission of the most requested contents is allocated orthogonal radio resources to avoid Inter-Cell Interference (ICI), and macro-diversity combining with optimal HM parameters is utilized to enhance the coverage area by reducing user outage. 

Hyper-local services transmission, on the other hand, deploys Hierarchical Modulation (HM) without any macro-diversity combining to transmit multimedia content within the scope of a cell, which can either be single-resolution or multi-resolution based on the distribution of the users requesting that content, and their radio link channel quality (CQI), obtained through an uplink feedback channel. 

\subsection{Multi-resolution Video Layering}
\begin{figure}[!t]
	\centering
	\includegraphics[width=3.5in]{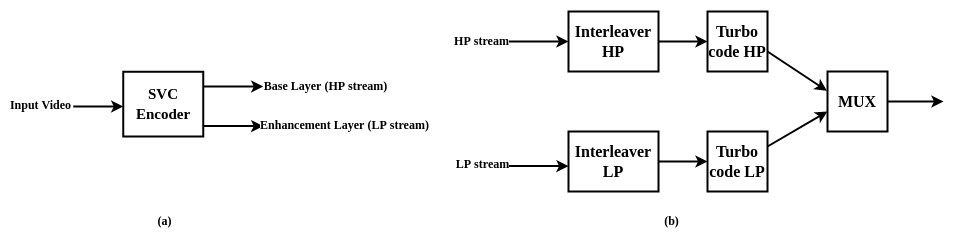}
	\caption{(a) Multi-resolution Source Coder and (b) Multi-resolution Channel Coder for LSA-wide/HLSA-wide Transmission}
	\label{label:mrTxChain}
\end{figure}

Visual data consists of layers of unequal importance in the perceived quality contribution at the receiver. For example, viewers in a live video conferencing scenario always pay more attention to the participants than the background scene. Out of the three main types of scalability offered by Scalable Video Coding (SVC), which is also an extension of H.264/AVC\cite{6121269}, the proposed LHS transmission scheme makes use of Quality scalability to decompose the input video stream into layers of different resolution levels, namely, Base and Enhancement Layers, as shown in Fig.[\ref{label:mrTxChain}]. 

\begin{figure}[!t]
	\centering
	\includegraphics[width=2.5in]{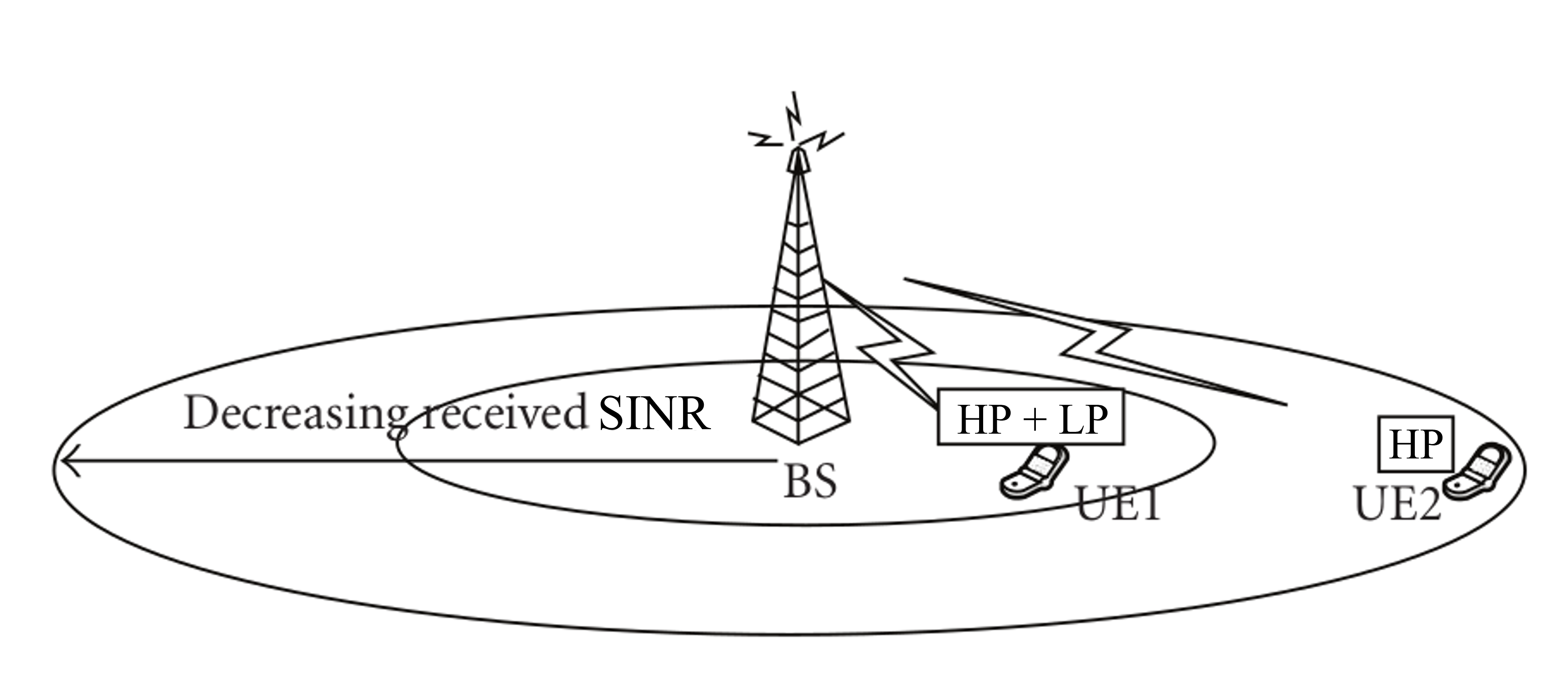}
	\caption{A broadcast model with HM.}
	\label{label:HM broadcast}
\end{figure}

Rather than sharing radio resources during the transmission of different video layers or resorting to Layered Division Multiplexing (LDM), we adopt the concept of Hierarchical Modulation (HM), as illustrated in Fig.[\ref{label:HM broadcast}] and elaborated in the subsequent section.

\subsection{Hierarchical modulation}
\begin{figure}[!t]
	\centering
	\includegraphics[width=2in]{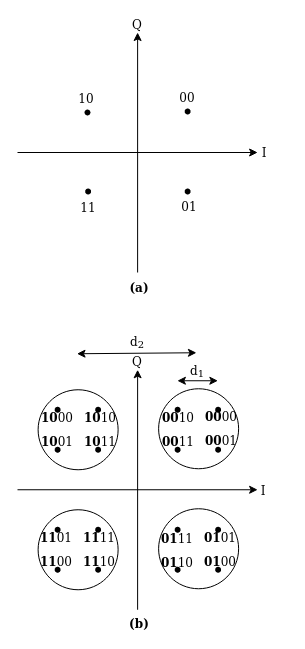}
	\caption{(a) QPSK constellation; (b) 16QAM hierarchical constellation. The bits in bold face are High Priority bits, and the rest are Low Priority bits.}
	\label{label:HM}
\end{figure}

 Broadcast systems such as Digital Video Broadcast Terrestrial (DVB-T), MediaFLO, and DVB-SH utilize Hierarchical Modulation (HM) to accommodate varying receiving conditions. In simple terms, HM can be visualized as a constellation comprising "clouds" of mini constellations or "satellites," where Low Priority (LP) information is represented in the satellites, and High Priority (HP) information is carried in the clouds\cite{210540}. For instance, the Hierarchically Modulated 16 QAM constellation in Fig.[\ref{label:HM}], with four bits per symbol, can be understood as four clouds, each containing four satellites. The HP stream, covering the entire service area, employs QPSK modulation, while the LP stream necessitates demodulating the constellation as 16 QAM. The LP bits may be used to transmit an additional Multimedia Service or an enhancement layer that enhances the resolution of the base layer transmitted via HP bits. Irrespective of the channel condition, a user always attempts to demodulate both the High Priority and Low Priority data streams\cite{Correia2009}.

\begin{figure}[!t]
	\centering
	\includegraphics[width=2.5in]{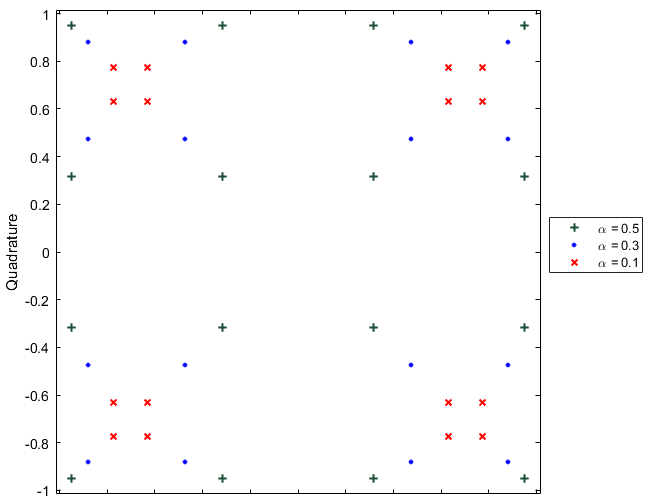}
	\caption{16QAM hierarchical constellation with $\alpha$ value set to 0.5, 0.3 and 0.1.}
	\label{label:constellation_comp}
\end{figure}

The amount of distortion that the LP symbols add to the HP constellation can be controlled by a design parameter that determines the constellation layout and is also used to control the ratio of coverage areas or service data rates\cite{1433080}.
    \[\frac{d_1}{d_2}= \alpha \quad   where \quad 0 < \alpha \leq0.5 \]
The 16 QAM constellation for different $\alpha$ values are as shown in Fig.\ref{label:constellation_comp},

To address capacity loss from Inter-Layer Interference (ILI),\cite{4536685} proposes an improved Hierarchical Modulation (HM) scheme called "HM with CR," which involves a counter-clockwise rotation of the Enhancement Layer (EL) signal constellation, optimizing the angle to maximize the minimum Euclidean distance (MED) between constellation points.

\subsection{OFDMA Resource Allocation}
\begin{figure}[!t]
	\centering
	\includegraphics[width=3.5in]{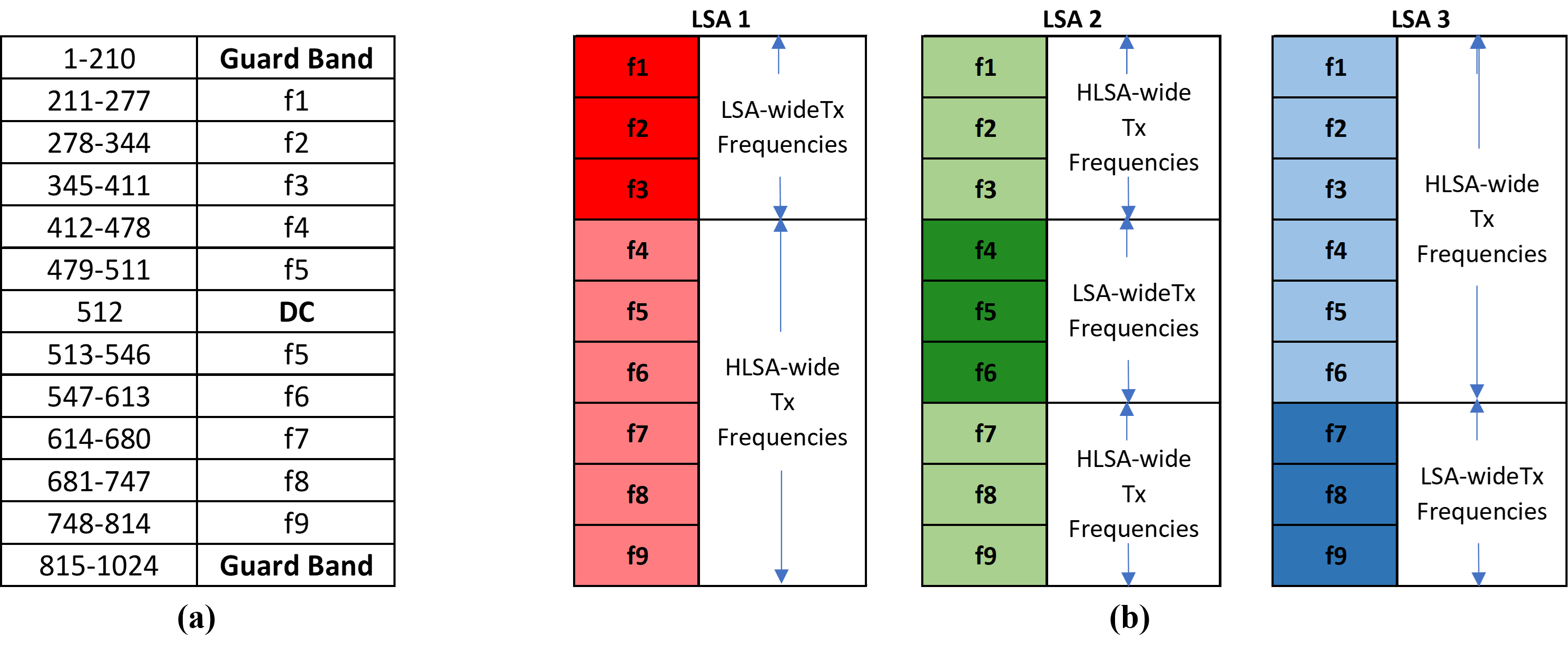}
	\caption{(a) Subcarrier Assignment and (b) Resource Allocation for the proposed design.}
	\label{label:res_alloc}
\end{figure}

The proposed system considers an OFDM symbol with 1024 subcarriers, of which 604 are utilized, including the DC signal. Each Multimedia Content request is allocated 67 subcarriers, accommodating 9 different Multimedia Content requests per OFDM symbol. As depicted in fig.[\ref{label:res_alloc}(a)], each subcarrier subband $f_i, i \in \{1, 2, ..., 9\}$ is dedicated to transmitting distinct Local Service Area (LSA)-wide or Hyper-Local Service Area (HLSA)-wide Multimedia content.

\begin{figure}[!t]
	\centering
	\includegraphics[width=1.5in]{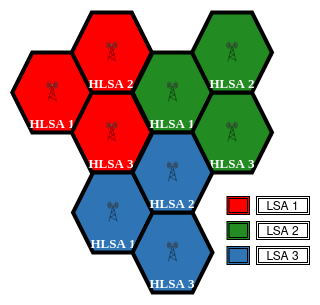}
	\caption{Local and Hyper-Local Service Areas in the proposed LHS transmission scheme}
	\label{label:LSA}
\end{figure}

The cluster of 9 cells is divided into 3 Local Service Areas (LSAs), each with 3 Hyper-Local Service Areas (HLSAs) whose coverage spans one cell, as depicted in Fig.[\ref{label:LSA}]. Each LSA is assigned an orthogonal sub-band of subcarriers for LSA-wide Multimedia content transmission within an OFDM Symbol which minimizes strong interference resulting from macro-diversity combining in nearby LSAs, consequently reducing service outage, as shown in Fig.[\ref{label:res_alloc}(b)]. However, this approach introduces strong interference to hyper-local content transmitted in adjacent cells, exceeding the interference experienced by typical Single Cell-Point to Multipoint (SC-PtM) users.

\subsection{UE Sub-grouping}
\begin{figure}[!t]
\centering
\includegraphics[width=0.8\linewidth]{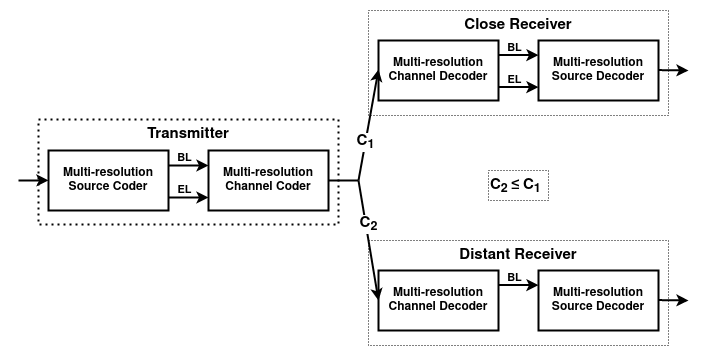}
\caption{Scalable video transmission.}
\label{label:svc}
\end{figure}

Implementing the Multicast Subgroup Formation concept\cite{6121269}, users in each cell requesting a specific multimedia content are subgrouped based on a threshold Channel Quality Indicator (CQI) value into two groups: High Channel Gain (HCG) Group and Least Channel Gain (LCG) Group. HCG Group users can decode both High Priority (HP) and Low Priority (LP) data streams, while LCG Group users may only decode the HP data stream, depending on their channel capacities, as depicted in Fig.[\ref{label:svc}].

The CQI threshold value was determined empirically by averaging the pre-processing received Signal-to-Noise Ratio (SNR) over each sub-band for $1\%$ Block Error Rate (BLER) measured for the $\alpha = 0.1$ Low Priority (LP) stream.

Using this knowledge of multicast group user distribution within each cell of the Local Service Area (LSA), we make decisions regarding the resolution of the transmitted multimedia content and whether a specific multimedia content request will be treated as a local service or a hyper-local service.

\subsection{Multimedia Content Request Mapping}
From the numerous multimedia content requests made within a Local Service Area (LSA), the selection of content for LSA-wide and HLSA-wide transmission on the allocated frequencies will be determined by the algorithms [\ref{LSA-chan-req}] and [\ref{HLSA-chan-req}], respectively.

\begin{algorithm}
    \caption{To find the 3 most requested multimedia contents for LSA-wide transmission.}
    \label{LSA-chan-req}
    \SetKwInOut{Input}{Input}
    \SetKwInOut{Output}{Result}
    \Input{Multimedia content request from all 3 cells within an LSA each with N dropped users  $r^i = [r_1^i , r_2^i, ..., r_N^i]$ where ${r_j^i} \in \{1, 2, ..., 15\}$, $ i \in \{1, 2, 3\}$ and  $ j \in \{1, 2,..., N\}$}
      \Output{3 multimedia contents for LSA wide transmission: \textit{lsa$\textunderscore$req}}
    \BlankLine
    $\%$Count the number of users requesting a Multimedia content within each cell of an LSA\\ 
   \For{i=1:3}{%
     \For{j=1:15}{%
     count$(i,j)$ = $\#$ users requesting Multimedia content $j$\;
     count$\textunderscore$hcg$(i,j)$ = $\#$ users requesting Multimedia content $j$ with received SNR $\geq$ SNR$\textunderscore$threshold\;
            }
            }
    total$\textunderscore$count = count$(1,:)$+ count$(2,:)$+ count$(3,:)$\;
    Sort total$\textunderscore$count in decreasing order and store the corresponding indices in sorted$\textunderscore$TC$\textunderscore$inds\\
    Initialise mm$\textunderscore$flag(\textit{j}) = 0, $j \in \{1, 2,..., 15\}$\\
    \For{i=1:3}{%
         lsa$\textunderscore$req$(i)$ = sorted$\textunderscore$TC$\textunderscore$inds$(i)$\;
         mm$\textunderscore$flag(\textit{lsa$\textunderscore$req}$(i)$) = 1 \;
        }
\end{algorithm}

\begin{algorithm}
    \caption{To find the multimedia contents to be transmitted on the HLSA-wide transmission frequencies in each cell within the LSA.}
    \label{HLSA-chan-req}
    \SetKwInOut{Input}{Input}
    \SetKwInOut{Output}{Result}
    \Input{Number of HCG group users requesting each multimedia content from all 3 cells within an LSA: \textit{count$\textunderscore$hcg}, Total number of users requesting each multimedia content from all 3 cells within an LSA: \textit{count}, multimedia flag indicating the multimedia contents selected for LSA-wide transmission: \textit{mm$\textunderscore$flag}.}
      \Output{Multimedia contents for HLSA wide transmission in each cell within the LSA - multi-resolution transmitted contents: \textit{hlsa$\textunderscore$mr$\textunderscore$req}, single-resolution transmitted contents both on LP: \textit{hlsa$\textunderscore$sr$\textunderscore$LP$\textunderscore$req} and HP bits: \textit{hlsa$\textunderscore$sr$\textunderscore$HP$\textunderscore$req}.}
    \BlankLine
     \For{i=1:3}{%
         Sort count$\textunderscore$hcg$(i,:)$ in decreasing order and store the corresponding indices in s$\textunderscore$idx\;
         hlsa$\textunderscore$mm$\textunderscore$flag = mm$\textunderscore$flag\;
         Set ctr = 1\;
         \For{j=1:15}{%
         \If{hlsa$\textunderscore$mm$\textunderscore$flag(s$\textunderscore$idx(j)) $\neq$ 1 $\&\&$ ctr $\leq$ 6} {%
                 \If{count$\textunderscore$hcg(i,s$\textunderscore$idx(j)) $\geq$ hcg$\textunderscore$threshold} {%
                    \uIf{count(i,s$\textunderscore$idx(j)) $>$ Total$\textunderscore$count$\textunderscore$threshold} {%
                        hlsa$\textunderscore$mr$\textunderscore$req$(i,ctr)$ = s$\textunderscore$idx$(j)$\;
                        hlsa$\textunderscore$mm$\textunderscore$flag(s$\textunderscore$idx$(j)$) = 1\;
                        Increment ctr\;
                    }
                   \Else{
                        hlsa$\textunderscore$mr$\textunderscore$req$(i,ctr)$ = -1\;
                        hlsa$\textunderscore$sr$\textunderscore$LP$\textunderscore$req$(i,ctr)$ = s$\textunderscore$idx$(j)$\;
                        hlsa$\textunderscore$mm$\textunderscore$flag(s$\textunderscore$idx$(j)$) = 1\;
                        sr$\textunderscore$flag = 1\;
                        \While {sr$\textunderscore$flag == 1 }{
                            \For{k=j:15}{%
                                \If{hlsa$\textunderscore$mm$\textunderscore$flag(s$\textunderscore$idx(k)) $\neq$ 1 $\&\&$ count$\textunderscore$hcg(i,s$\textunderscore$idx(k)) $<$ hcg$\textunderscore$threshold $\&\&$ count(i,s$\textunderscore$idx(k)) $>$ Total$\textunderscore$count$\textunderscore$threshold} {%
                                    hlsa$\textunderscore$sr$\textunderscore$HP$\textunderscore$req$(i,ctr)$ = s$\textunderscore$idx$(k)$\;
                                    hlsa$\textunderscore$mm$\textunderscore$flag(s$\textunderscore$idx$(k)$) = 1\;
                                    Increment ctr\;
                                    sr$\textunderscore$flag = 0\;
                                    break\;
                                }
                            }
                        }
                    }
                }   
            }
        }
    }   
\end{algorithm}
Therefore, the three most requested multimedia contents will be transmitted on the LSA-wide transmission frequencies allocated to that LSA. The remaining multimedia content requests within a Hyper-Local Service Area (HLSA) will be transmitted on the HLSA-wide transmission frequencies and can be either single or multi-resolution based on user distribution. The scheduling algorithm [\ref{HLSA-chan-req}] is succinctly explained in the Table[\ref{label:UEtable}],

\begin{table}[!t]
\caption{Multimedia Content Request Mapping Logic.\label{label:UEtable}}
\centering
\resizebox{\columnwidth}{!}{%
\begin{tabular}{l||c|c|c|c}\hline
\multirow{3}{*}{} & \multirow{3}{*}{\textbf{LSA-wide Transmission}} &  \multicolumn{3}{c}{\textbf{HSLA-wide Transmission}} \\
\cline{3-5}
 & & \multirow{3}{*}{\textbf{Multi-Resolution}} & \multicolumn{2}{c}{\textbf{Single-Resolution}}\\\cline{4-5}
 & & & \textbf{HP} & \textbf{LP}
\\\hline\hline
\textbf{Criteria} &1&2&3&4\\\hline
\textbf{$\%$ HCG users} &&$\geq50\%$&$<50\%$&$\geq50\%$\\\hline
\multirow{4}{*}{\textbf{Total \# users}} &{Total \# users in cell 1 + }&\multirow{4}{*}{$>$Total$\textunderscore$count$\textunderscore$threshold}&\multirow{4}{*}{$>$ Total$\textunderscore$ count$\textunderscore$threshold}&\multirow{4}{*}{$\leq$ Total$\textunderscore$count$\textunderscore$threshold}\\
&{Total \# users in cell 2 +}&&\\
&{Total \# users in cell3}&&\\
&{= Highest in the LSA}&&
\\\hline
\end{tabular}%
}
\end{table}

For Single Cell-Point to Multipoint (SCPtM), multicast groups are scheduled based on criteria 1, 2, and 3. The modulation order is 16 QAM if there is not even one LCG user; otherwise, it is QPSK.

\begin{figure}[!t]
	\centering
	\includegraphics[width=2.7in]{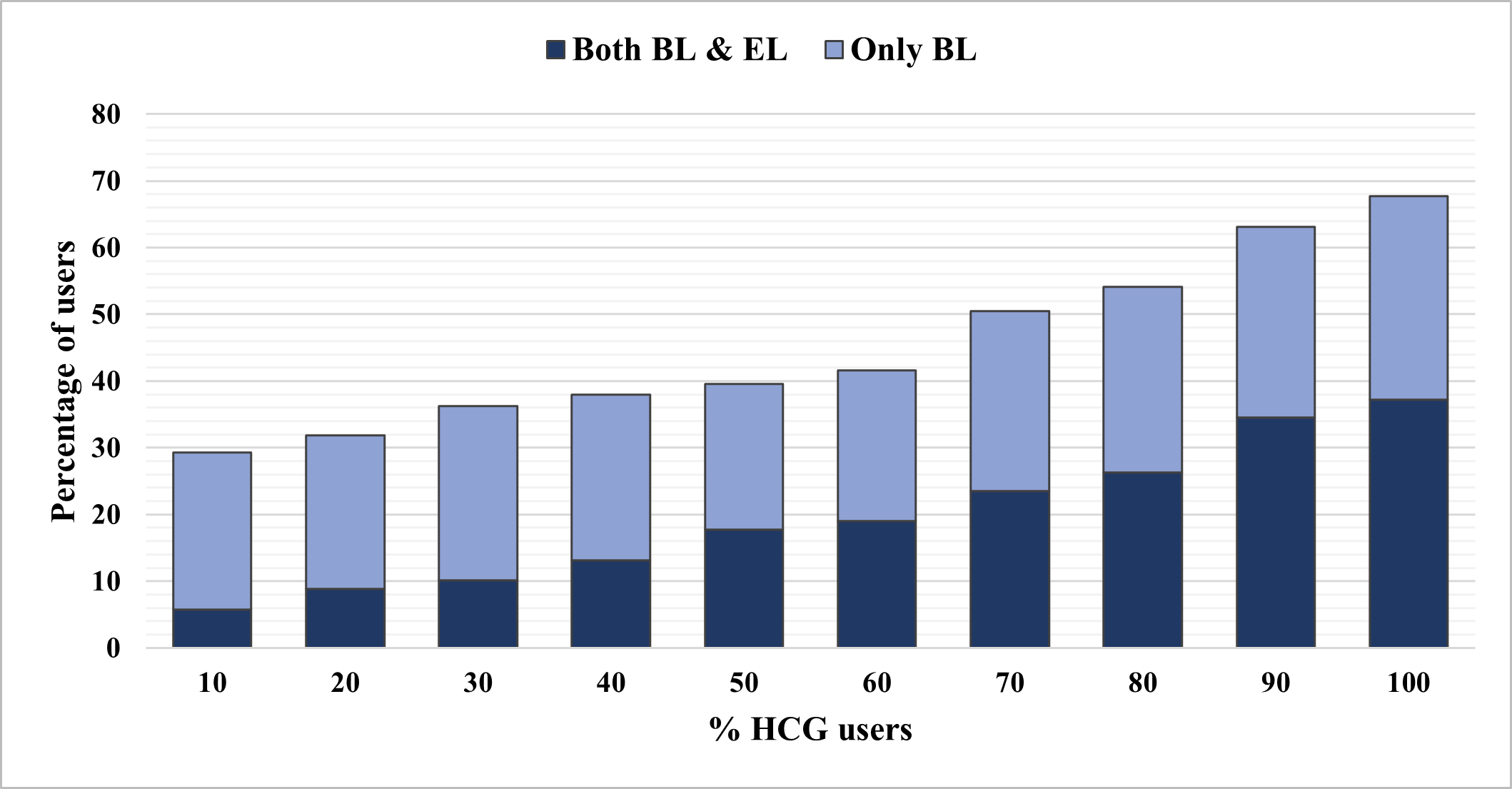}
	\caption{Percentage of users who can successfully decode each layer vs $\%$ High Channel Gain (HCG) users.}
	\label{label:outage}
\end{figure}

As shown in Fig.[\ref{label:outage}], selecting a percentage of High Channel Gain (HCG) users to be above 50$\%$ is crucial because only when this threshold is surpassed can more than 20$\%$ of total users successfully decode the Enhancement Layer (EL). Transmitting EL has no practical benefit if only a few users can successfully decode it.

For subgroups satisfying Criteria 3, which consists of more Least Channel Gain (LCG) users, SCPtM transmits single-resolution multimedia content using QPSK, equivalent to setting the Hierarchical Modulation (HM) parameter $\alpha$ = 0. These users are likely closer to the cell edge, making them susceptible to Inter-Cell Interference (ICI) and thus prone to outage. Instead of providing additional protection to these users by switching to a 4 QAM-only system, it is more appropriate to offer an extra service to multicast groups satisfying Criteria 4, which has more High Channel Gain (HCG) users capable of receiving service through Low Priority (LP) bits. This multicast group is less prone to ICI, which ensures less outage.

The proposed design sets $\alpha$ = 0.3 to multiplex these two services, shifting protection from the High Priority (HP) layer to the Low Priority (LP) layer to provide equal protection to both. This results in reduced coverage of the HP layer compared to the original 4 QAM constellation. The introduced Inter-Layer Interference (ILI) can be effectively handled using the described technique\cite{4536685}.

Thus, an additional service is provided at the cost of reduced coverage for one multimedia content service, leading to an overall increase in the total number of multimedia content services offered per Hyper-Local Service Area (HLSA). 

\subsection{LSA-wide Multimedia Content Transmission}
Unlike MBSFN, which provides reserved cell space around its transmission area and cannot be reduced to a three-cell service area due to excessive spectrum wastage, the proposed Local Service Area (LSA) design exposes the cell-edge users to Inter-Cell Interference (ICI) since no provision is made for reserved cells.

The proposed LHS transmission scheme applies non-uniform data protection across the entire video stream. The Base Layer (BL) is given more protection than the Enhancement Layer (EL) by using Hierarchical Modulation (HM) and macro-diversity combining since the EL information can only be utilized if all the BL information is received correctly\cite{1618934}.

\begin{figure}[!t]
	\centering
	\includegraphics[width=1.35in]{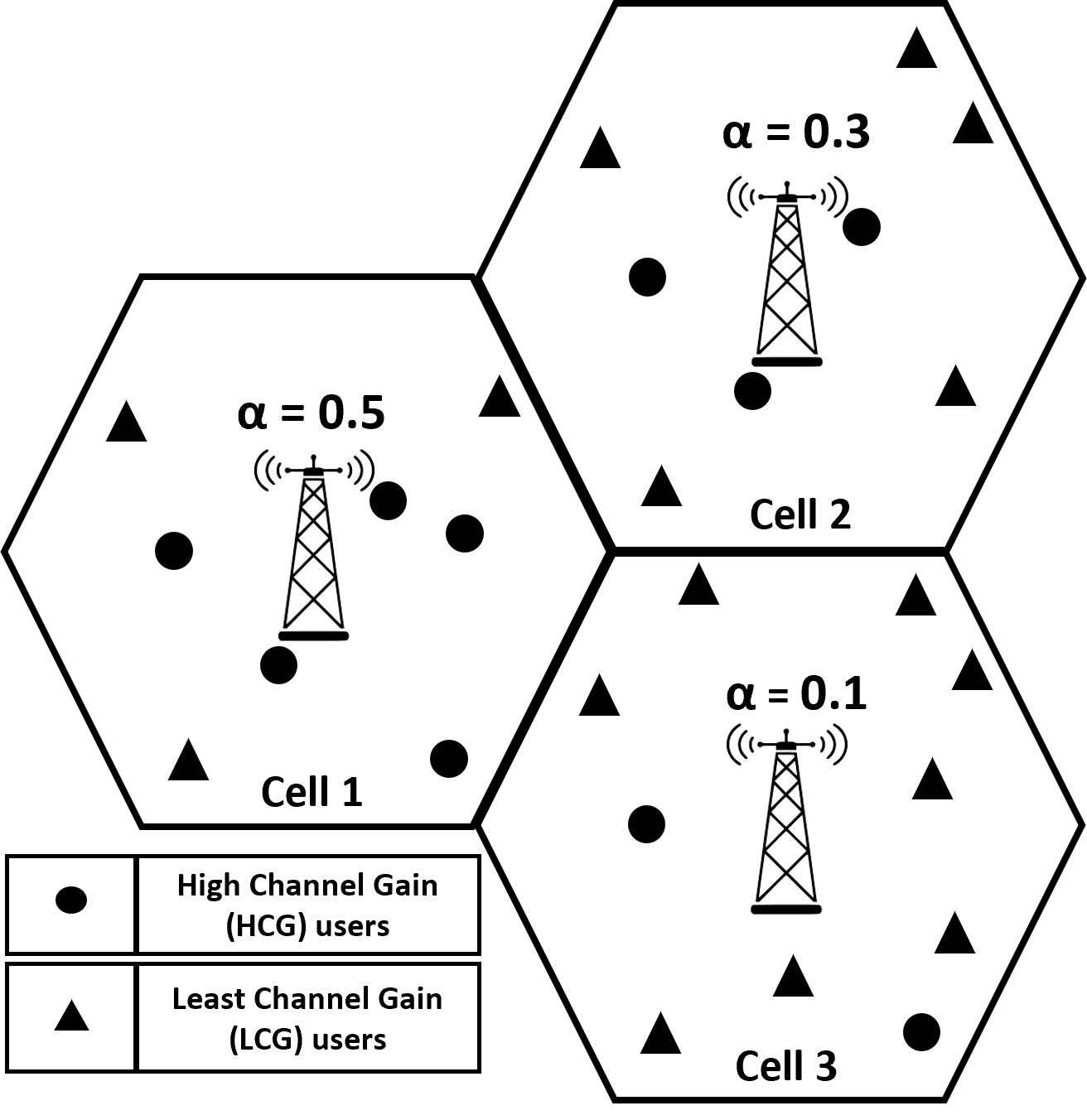}
	\caption{Dynamic $\alpha$ assignment}
	\label{label:alphaAss}
\end{figure}

Macro-diversity combining prevents ICI from adjacent cells while providing diversity gain. The three cells within an LSA collaboratively transmit the same multimedia content to the group of users requesting that service. Authors in\cite{Correia2009} suggest assigning HM parameter $\alpha$ = 0.5 to each link. However, the proposed LHS transmission scheme adaptively assigns $\alpha$ values to these links based on the number of High Channel Gain (HCG) and Least Channel Gain (LCG) users in each cell within an LSA, as depicted in Fig.[\ref{label:alphaAss}] and described in algorithm[\ref{alphaAss}]. The dynamic assignment of $\alpha$ values facilitates reliable reception of BL data for users in each cell, even if the links from the other two gNBs are in a deep fade, at the expense of reduced diversity gain for EL.

\begin{figure}[!t]
	\centering
	\includegraphics[width=3.9in]{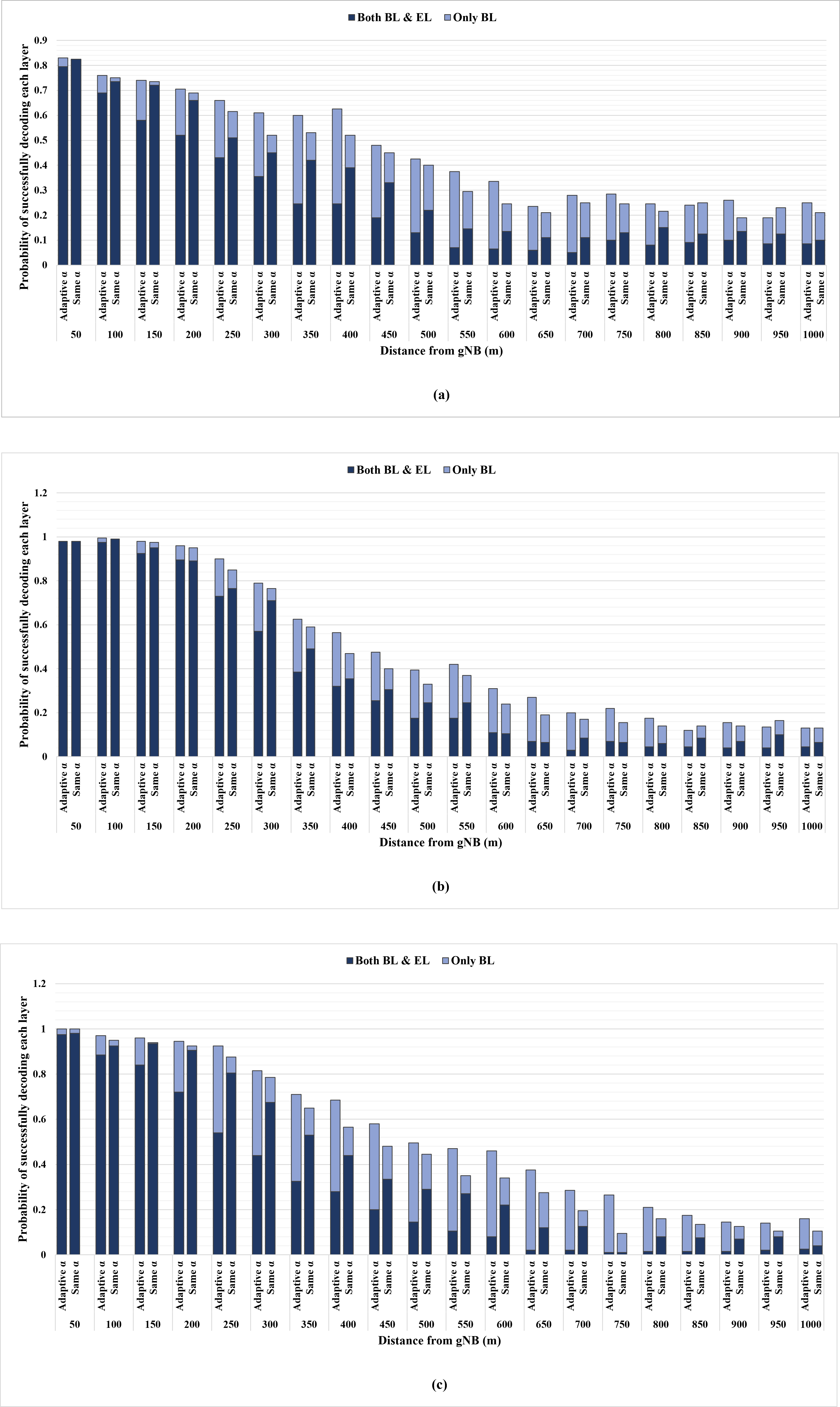}
	\caption{Probability of successfully decoding each layer vs distance from gNB (a) Cell 0 (b) Cell 1 (c) Cell 2}
	\label{label:avgThru}
\end{figure}

Fig. [\ref{label:avgThru}] illustrates the probability of successfully decoding each layer as the user's distance from the gNB increases within each cell of the Local Service Area(LSA). The performance of our proposed adaptive $\alpha$ assignment scheme is compared to the scheme in\cite{Correia2009}, denoted as "same alpha" in the plots. The graphs indicate a slightly higher probability for users to successfully decode both the Base Layer (BL) and Enhancement Layer (EL) in each cell with the "same alpha" scheme. However, this degradation is counterbalanced by a corresponding increase in the probability for the user to successfully decode only BL. Thus, there is a trade-off between group satisfaction and fairness, ultimately leading to a reduction in the number of users experiencing service outage in each cell of the Local Service Area (LSA). Also, as the user's distance from the gNB increases, the probability of successfully decoding each layer decreases, regardless of the $\alpha$ assignment scheme, owing to Inter-Cell Interference (ICI) from neighboring cells.

\begin{algorithm}
    \caption{To assign $\alpha$ values to all 3 cells of an LSA for LSA-wide multimedia content transmission.}
    \label{alphaAss}
    \SetKwInOut{Input}{Input}
    \SetKwInOut{Output}{Result}
    \Input{Number of HCG and LCG group users requesting the LSA-wide multimedia content in all 3 cells of an LSA}
      \Output{$\alpha$ values assigned to all 3 cells of an LSA for a particular multimedia content to be transmitted LSA-wide.}
    \BlankLine
    Initialize $\alpha$ = [0.5 0.3 0.1]\;
    $\%$ Define counters C and TC  \\
    \For{i=1:3}{%
         C$(i)$ = $\#$ HCG group users in cell $i$\;
         TC$(i)$ = $\#$ HCG group users + $\#$ LCG group users in cell $i$\;
            }
    Sort C in descending order and store it in sorted$\textunderscore$C and its corresponding indices in sCinds \\
    Sort TC in descending order and store it in sorted$\textunderscore$TC and its corresponding indices in sTCinds  \\     
    \uIf{sorted$\textunderscore$C has all unique elements} {
         $\alpha[$sCinds(1) sCinds(2) sCinds(3)]  = [0.5 0.3 0.1]\;
    }
    \Else{
        \uIf{sorted$\textunderscore$C has all equal elements} {
            \uIf{sorted$\textunderscore$TC has all unique elements} {
                $\alpha[$sTCinds(1) sTCinds(2) sTCinds(3)]  = [0.5 0.3 0.1]\;     
            }
            \Else{
                \uIf{sorted$\textunderscore$TC has all equal elements} {
                 $\alpha$ remains unchanged
                }
                \Else{
                    \uIf{sorted$\textunderscore$TC(1) == sorted$\textunderscore$TC(2) }{
                        $\alpha[$sTCinds(1) sTCinds(2) sTCinds(3)]  = [0.5 0.3 0.1]\;     
                    }
                    \Else{
                        $\alpha[$sTCinds(1) sTCinds(2) sTCinds(3)]  = [0.5 0.3 0.1]\;     
                    }
                }
            }
        }
        \Else{
            \uIf{sorted$\textunderscore$C(1) == sorted$\textunderscore$C(2) }{
                \uIf{TC[sCinds(1)] $\geq$ TC[sCinds(2)]]}{
                    $\alpha[$sCinds(1) sCinds(2) sCinds(3)]  = [0.5 0.3 0.1]\;
                }
                \Else{
                    $\alpha[$sCinds(1) sCinds(2) sCinds(3)]  = [0.3 0.5 0.1]\;
                }
            }
            \Else{
                \uIf{TC[sCinds(2)] $\geq$ TC[sCinds(3)]]}{
                    $\alpha[$sCinds(1) sCinds(2) sCinds(3)]  = [0.5 0.3 0.1]\;
                }
                \Else{
                    $\alpha[$sCinds(1) sCinds(2) sCinds(3)]  = [0.5 0.1 0.3]\;
                }
            }
        }
    }
\end{algorithm}

\subsection{HLSA-wide Multimedia Content Transmission}
\begin{figure}[!t]
	\centering
	\includegraphics[width=2.5in]{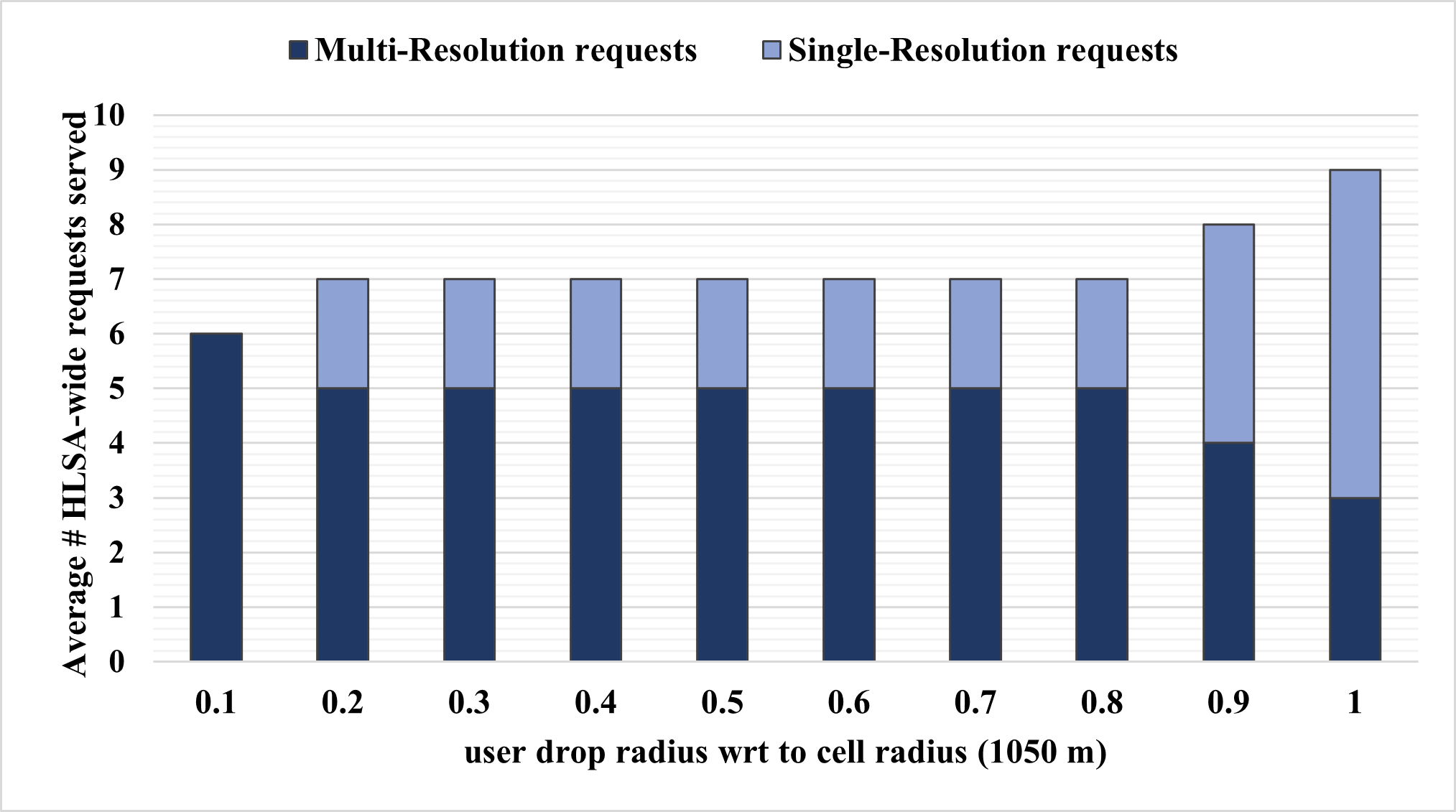}
	\caption{Average number of Hyper-Local Service requests served per cell. The users are dropped within the user drop radius defined with respect to the cell radius.}
	\label{label:hlsareqs}
\end{figure}

No macro-diversity combining is deployed for HLSA-wide multimedia content transmission, and the reception area is limited to a single cell. The HLSs being scheduled will have a lesser number of users requesting them as compared to user strength requesting the LSA-wide multimedia contents. Algorithm [\ref{HLSA-chan-req}] describes a scheduling procedure where the Low Priority bits of Hierarchical Modulation may be used to transport an additional Multimedia Content service or an enhancement stream that enhances the resolution of the base stream based on the users' CQI profile. Thus, the proposed LHS transmission scheme can support at most 6 multi-resolution Multimedia Content services or 12 single-resolution Multimedia Content services for HLSA-wide transmission.
The figure [\ref{label:hlsareqs}] illustrates that, on average, the proposed LHS transmission scheme can support more Hyper-Local Services (HLSs) compared to Single Cell-Point to Multipoint (SCPtM). 

\begin{figure}[!t]
	\centering
	\includegraphics[width=3.5in]{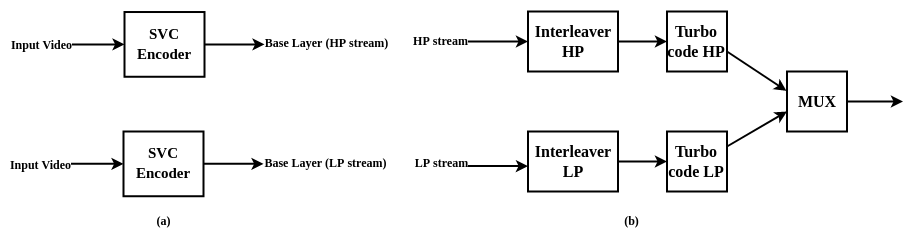}
	\caption{(a) Single-resolution Source Coder and (b.) Single-resolution Channel Coder for HLSA-wide Transmission.}
	\label{label:srTxChain}
\end{figure}

For single-resolution content transmission, the data loaded onto the LP and HP bits will be single-resolution base layer datastream output from the SVC encoder for 2 different input video sources as shown in Fig.[\ref{label:srTxChain}]. Whereas, for the multi-resolution content transmission, the source and channel coder will be the same as that used for LSA-wide Transmission in Fig.[\ref{label:mrTxChain}].

\begin{figure*}[!t]
	\centering
	\includegraphics[width=\textwidth]{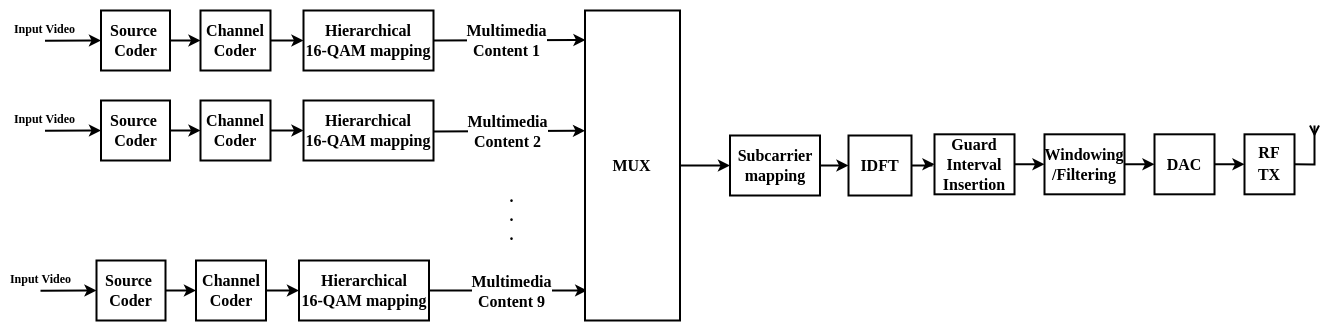}
	\caption{Transmitter block diagram of the proposed LHS transmission scheme.}
	\label{label:tx Block diagram}
\end{figure*}

\subsection{Transmitter block diagram}
The transmitter block diagram for the proposed LHS transmission scheme is shown in Figure [\ref{label:tx Block diagram}]. Each multimedia content follows a distinct chain of source coder, channel coder, and HQAM mapper based on its transmission type (LSA-wide or HLSA-wide) or resolution (single or multi-resolution) before the data is multiplexed and loaded onto their respective bands of subcarriers.

\section{Performance Evaluation}

Radio network simulations are generally categorized into link level, which focuses on the radio link between the gNB and the user terminal, and system level, which involves multiple gNBs and mobile users. While having a unified approach would be ideal, the complexity of creating a simulator covering everything from transmitted waveforms to a multicell network is too high for the required simulation resolutions and time. Therefore, an interconnected setup of separate link and system-level simulators is necessary. 

The link level simulator is crucial for the system simulator to construct a receiver model predicting Block Error Rate(BLER) performance, accounting for factors like channel estimation, interleaving, modulation, receiver structure, and decoding. Meanwhile, the system-level simulator models a system with numerous mobiles and gNBs, along with the algorithms operating within such a system.

\subsection{Link-Level Simulator Design}
The link-level simulator (LLS) was developed in Matlab and followed the 3GPP Release specifications.The LLS encompassed the entire OFDMA signal processing at the transmitter, incorporating various receiver structures. The simulator considered Multipath Rayleigh fading channels due to the sensitivity of hierarchical high-order QAM modulations to channel variations. For simplicity, the simulations assumed perfect knowledge of channel gains at each receiver.

\subsection{Radio Access Network System Level Simulator}
A system-level simulator was developed in C++ using the IT++ library, an open-source tool for efficient matrix manipulation and signal processing in communication systems. This simulator, operating at a Transmission Time Interval (TTI) rate of 1 ms, captures the dynamic aspects of the Radio Access Network System, including user aspects like mobility and variable traffic demands, the Radio Access Network (RAN) with a certain level of abstraction, and the radio interface. Table [\ref{tab:system_parameters}] outlines the link and system-level simulation parameters derived from 3GPP documents \cite{3gpp1,3gpp2,3gpp3} .

\begin{table}[!t]
\caption{Link and System-level simulation parameters.\label{tab:system_parameters}}
\centering
\resizebox{\columnwidth}{!}{%
\begin{tabular}{|l|l|}
\hline
\textbf{Parameter} & \textbf{Assumption} \\\hline
\hline
Transmission Bandwidth & 10 MHz \\\hline
Carrier Frequency & 2120 MHz \\\hline
FFT Size & 1024 \\\hline
Sampling Frequency & 15.36 MHz \\\hline
Number of Used Subcarriers (including DC) & 604 \\\hline
Sub-carrier Spacing & 15 kHz \\\hline
Available Bandwidth & 9.060 MHz \\\hline
Cellular Layout & Hexagonal grid, 57 cell sites \\\hline
Inter-site Distance (m) & 2000 \\\hline
Path Loss Model Environment & ITU-R Urban Macro–URLLC Configuration A \\\hline
Penetration Loss (dB) & 0 \\\hline
Shadow Loss Variance (LOS/nLOS) & 8/12 dB \\\hline
Coherence Time & 1 ms \\\hline
Max Doppler Frequency & 423.0 Hz \\\hline
Coherence Bandwidth & 315.75 kHz \\\hline
gNB Transmit Power & 44 dBm \\\hline
gNB Antenna Gain & 8 dBi \\\hline
Noise Spectral Density & -174 dBm/Hz \\\hline
UE Antenna Gain & 0 dBi \\\hline
UE Noise Figure & 6 dB \\\hline
Short Term Channel Model (6 taps) & Ped-B \\\hline
Fast Fading Model & Jakes’ model \\\hline
CP Length (Long CP) (µs) & 16.67 \\\hline
BLER Target & 1\% \\\hline
TTI (Time Transmission Interval) & 1 ms \\\hline
Threshold for Total Users Requesting Multimedia Content & 10 \\
\hline
\end{tabular}%
}
\end{table}

The channel model in the system-level simulator incorporates three types of losses: path loss, shadowing loss, and multipath fading loss (one value per Transmission Time Interval). The model parameters are environment-dependent, with the path loss model based on the urban macro-mMTC test Configuration A from ITU-R guidelines \cite{series2017guidelines}. Shadowing, which results from large obstacles like buildings and UE movements, is modeled through a lognormal distribution with a correlation distance. The multipath fading follows the 3GPP channel model, using the ITU Pedestrian B environment as a reference whose correlated instances were generated using Jakes' Model.

In OFDM systems, the maximum delay spread of the multipath channel influences the length of the Cyclic Prefix chosen to prevent intersymbol interference. The proposed system adopts a long Cyclic Prefix (CP) to accommodate Time-of-Flight (TOF) variations from the gNB to the cell-edge user, at the cost of reduced transmitted bit rates.

We examine a scenario with 57 cells, a configuration commonly employed by 3GPP. Each gNB has a coverage radius of 1000 m. The gNB transmit power is 44 dBm and its antenna gain is 8 dBi and the antenna gain is 0 dBi for the User Equipment. 

\begin{figure}[!t]
	\centering
	\includegraphics[width=2in]{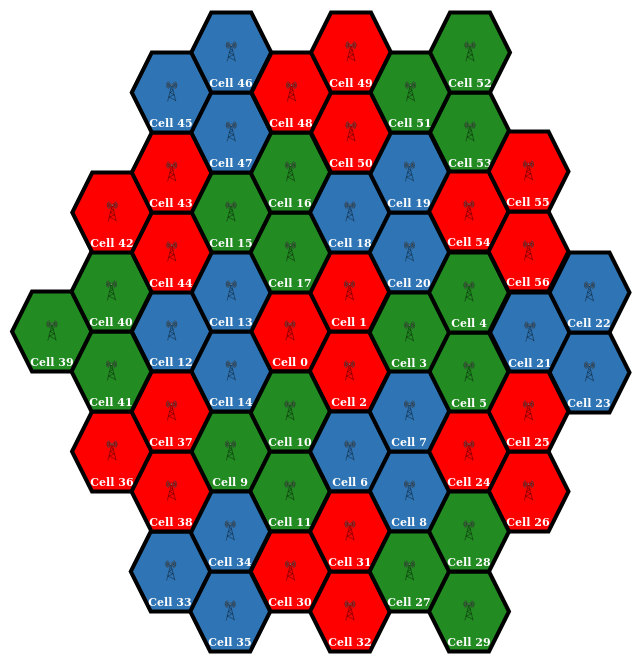}
	\caption{Cellular Layout for system-level simulations.}
	\label{label:cluster}
\end{figure}

At the start of each simulation, users are randomly distributed within each cell of the considered Local Service Area (LSA), and their positions are assumed to be fixed. The system-level simulator models interference from neighboring tier 1 and 2 cells surrounding each cell in the center Local Service Area (LSA). Fig.[\ref{label:cluster}] illustrates the proposed scheme's cellular layout, depicting the orthogonal frequency reuse pattern for the locally transmitted services across the cluster.

The minimum data rate for the Base and the Enhancement Layer is 800 Kbps. In the SCPtM system, the data rates for SD and HD are 800 Kbps and 1.604 Mbps, respectively.

The proposed LHS transmission scheme's performance is evaluated through simulations and compared to the SC-PtM scheme. The simulation scenario accounts for variations in the number of users requesting a specific multimedia service in each cell.

Multiple iterations of each simulation run were conducted to establish 95$\%$ confidence intervals for the key results.

The schemes were compared based on the following performance metrics:
\begin{itemize}
     \item \textbf{Percentage of users receiving the local service} measures the percentage of users experiencing different qualities(SD or HD) of a particular local service in each cell of the center Local Service Area (LSA).
    
    \item \textbf{Percentage of users receiving the multi-resolution hyper-local service} measures the percentage of users experiencing different qualities(SD or HD) of a particular hyper-local service in cell 0 of the center LSA.
    
    \item \textbf{Percentage of users receiving the single-resolution hyper-local service} measures the percentage of users receiving each hyper-local service of SD quality in cell 0 of the center LSA.
\end{itemize}

\subsection{Simulation Results}
\begin{figure}[!t]
	\centering
	\includegraphics[width=3in]{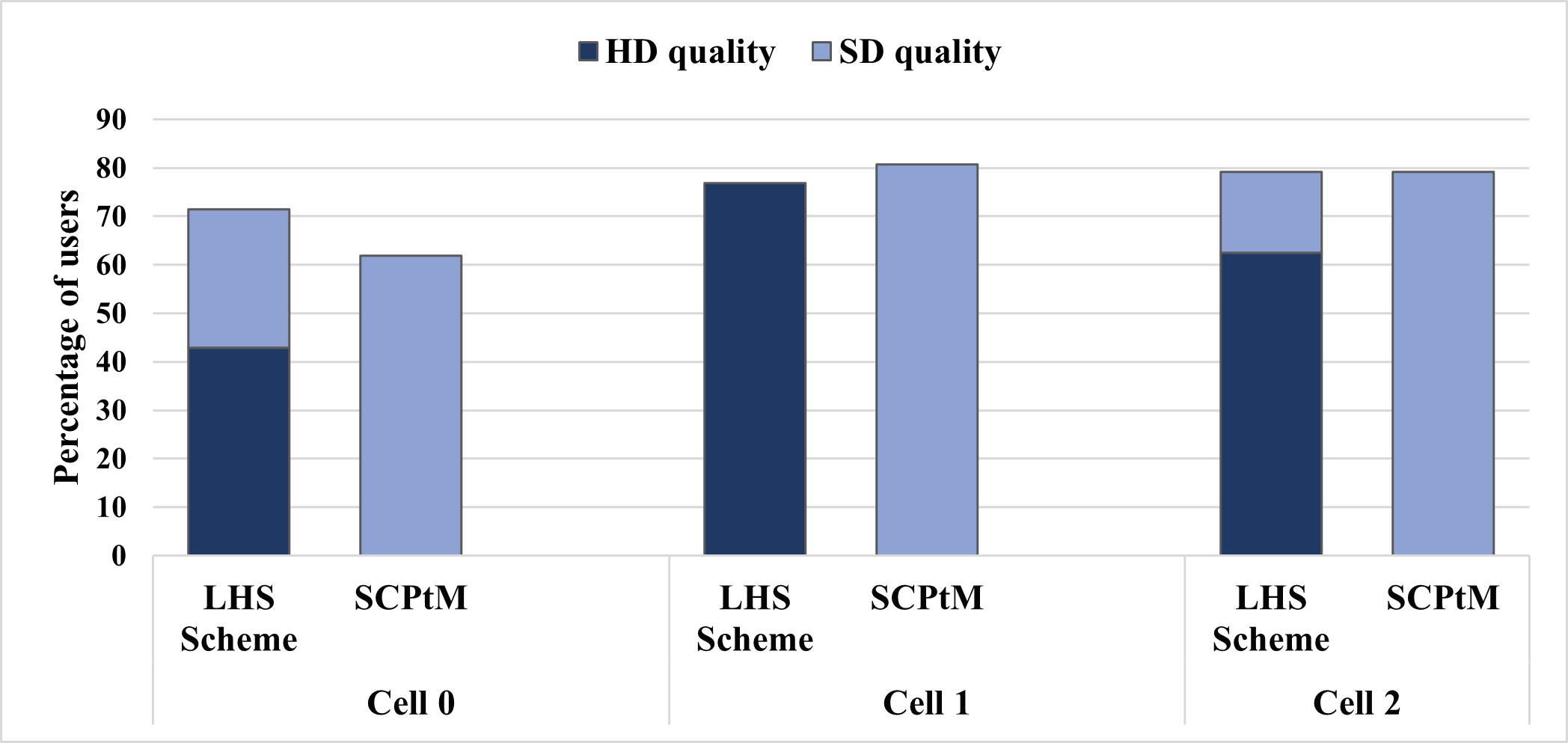}
	\caption{Percentage of users receiving the local service with $\alpha$ values set to 0.1, 0.5, and 0.3 in cells 0, 1, and 2, respectively.}
	\label{label:LS_op}
\end{figure}

The shaded portion of the bar graph in Fig. [\ref{label:LS_op}] illustrates the percentage of users experiencing HD-quality local service who can successfully decode both Base and Enhancement layers in the proposed scheme. Successful reception of the Base Layer is essential to utilize the Enhancement Layer. Hence, the users who cannot successfully decode the Base Layer, regardless of the Enhancement Layer's decoding status, are considered to be in service outage. The unshaded portion of the bar graph represents users experiencing SD-quality local service who can successfully decode only the Base Layer. 

The performance was evaluated in cells 0, 1, and 2 of the center Local Service Area (LSA). In each cell, SCPtM utilized QPSK for transmitting SD quality local service due to the LCG principle, whereas the proposed design employed macro-diversity combining with $\alpha$ values set to 0.1, 0.5, and 0.3 in cells 0, 1, and 2, respectively.

As a result of deploying hierarchical modulation, 43$\%$, 77$\%$, and 63$\%$ of the total dropped users in cells 0, 1, and 2, respectively, experience HD quality in the local service. Additionally, 29$\%$, 0$\%$, and 17$\%$ of the total dropped users in cells 0, 1, and 2, respectively, experience SD quality local service.

As expected, in the proposed LHS transmission scheme, fewer users experience service outage in each cell of the Local Service Area (LSA) compared to SCPtM due to microdiversity combining. However, some users still face outage due to Inter-Cell Interference (ICI) from adjacent cells. The utilization of hierarchical modulation improves the satisfaction of HCG users, enabling them to receive higher-quality service and increasing the multicast group throughput in the LSA.

The assigned $\alpha$ values influence the trend in users experiencing HD quality in the local service. Cell 1, with more HCG users, was assigned $\alpha$ = 0.5, followed by $\alpha$ = 0.3 in cell 2 and $\alpha$ = 0.1 in cell 0. A higher $\alpha$ value provides more protection to the Enhancement Layer (EL), resulting in more users experiencing HD quality in cell 1, followed by cell 2 and cell 0. However, this increased protection causes a slight decrease in the number of users experiencing SD quality in cell 1 due to Inter-Layer Interference (ILI) induced from Low Priority (LP) to High Priority (HP) bitstream.

\begin{figure}[!t]
	\centering
	\includegraphics[width=2.1in]{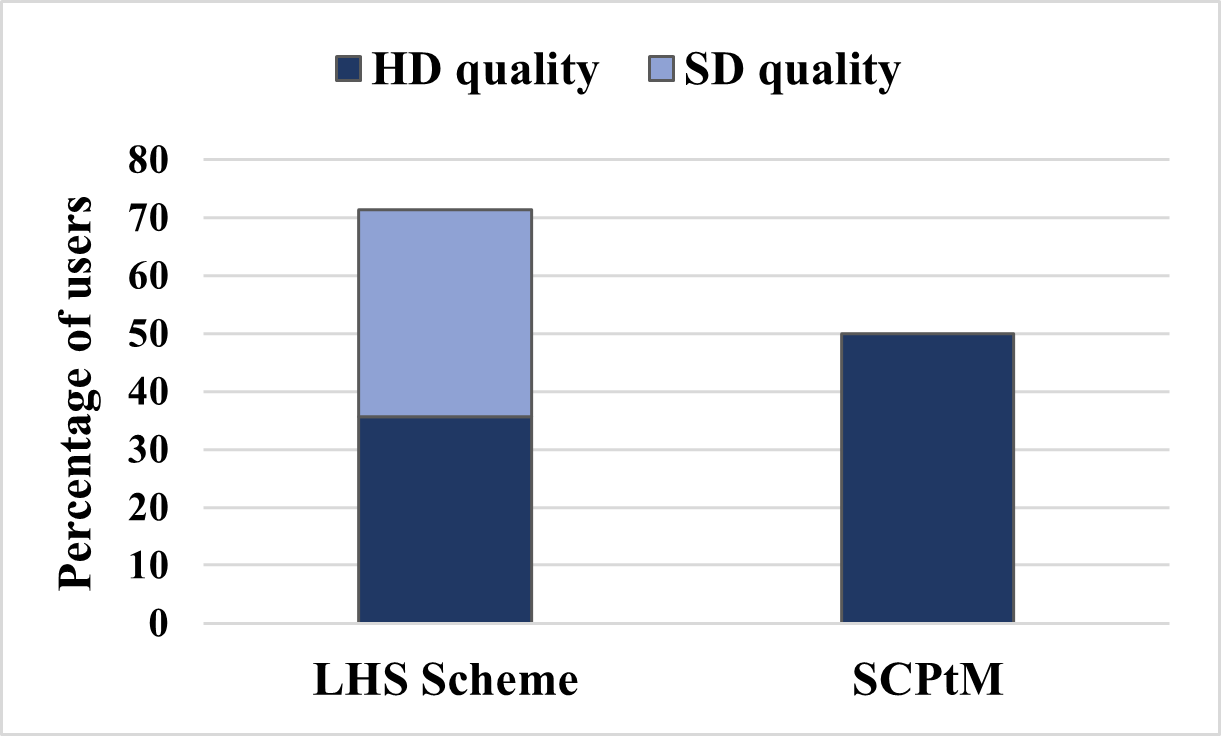}
	\caption{Percentage of users receiving the multi-resolution hyper-local service.}
	\label{label:MR_HLS}
\end{figure}

Fig. [\ref{label:MR_HLS}] illustrates the second performance metric, measuring the percentage of users experiencing HD quality in the hyper-local service,  which refers to those who successfully decode each layer when a multi-resolution hyper-local service is transmitted in cell 0 of the center LSA. In the absence of an LCG user, SCPtM utilized non-hierarchical 16 QAM and transmitted HD quality hyper-local service.

In the proposed LHS transmission scheme, 36$\%$ of the total dropped users in cell 0 experience HD quality hyper-local service, compared to 50$\%$ in the SCPtM. This decrease is again attributed to the impact of Inter-Layer Interference (ILI). However, it is counterbalanced by a corresponding decrease in users experiencing service outage, as an additional 36$\%$ of the total dropped users can experience SD quality hyper-local service in the proposed scheme.

\begin{figure}[!t]
	\centering
	\includegraphics[width=2in]{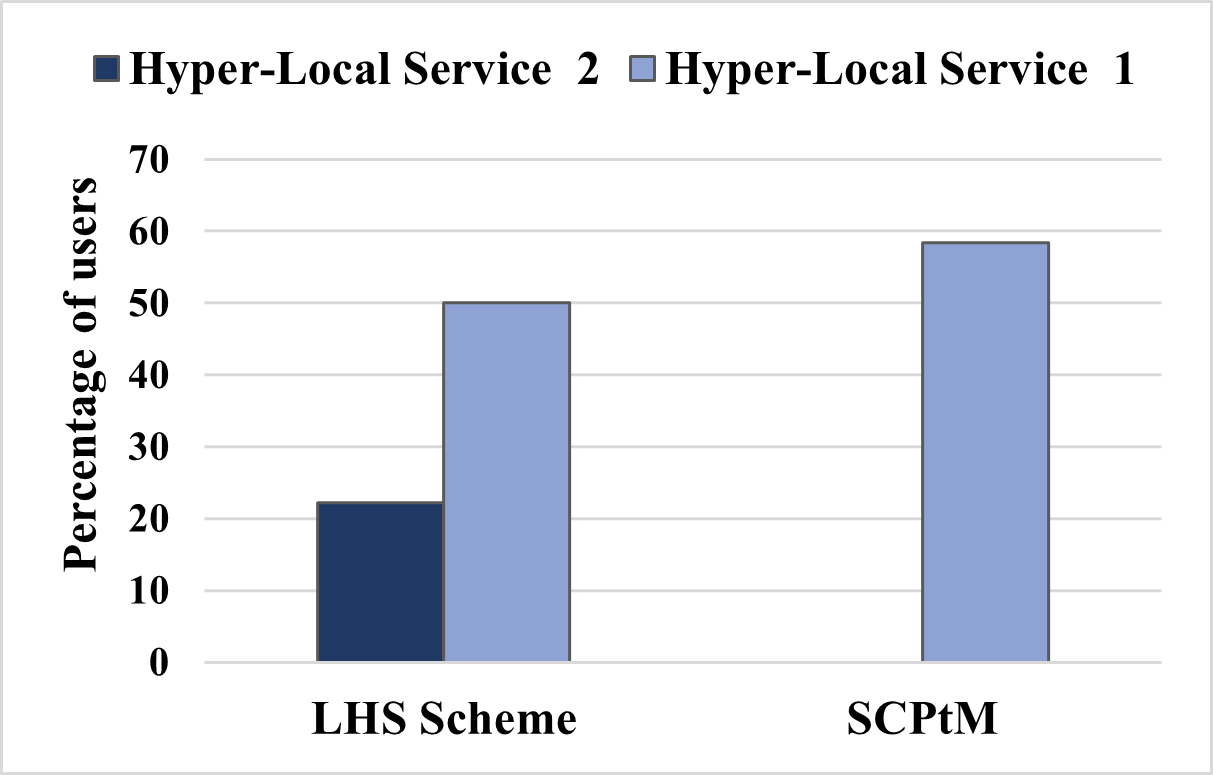}
	\caption{Percentage of users receiving the single-resolution hyper-local service.}
	\label{label:SR_HLS}
\end{figure}

The third performance metric, which measures the percentage of users receiving each hyper-local service of SD quality in cell 0 of the center  Local Service Area (LSA), is presented in Fig. [\ref{label:SR_HLS}]. The impact of Inter-Layer Interference (ILI) induced by the transmission of the Low Priority (LP) layer alongside the High Priority (HP) layer is evident, resulting in a slight decrease in the percentage of users able to receive hyper-local service 1—from 58$\%$ in SCPtM to 50$\%$ in the proposed LHS transmission scheme. However, this allows for serving approximately 22$\%$ of users requesting hyper-local service 2. The percentage is comparatively lower due to the chosen $\alpha$ value for single-resolution hyper-local service transmission. Selecting a higher $\alpha$ value could have facilitated serving a slightly larger number of users requesting hyper-local service 2, albeit at the cost of a proportional decrease in the percentage of users able to receive hyper-local service 1.

\section{Conclusion}
The paper introduced a novel Local and Hyper-Local Services (LHS) transmission scheme that improves the spectral efficiency of cellular networks by optimally allocating resources for each requested multimedia content based on user distribution. A comparison with the SCPtM transmission scheme revealed that: (i) Transmitting multi-resolution multimedia local and hyper-local services enhanced the Quality of Experience (QoE) for multicast group users. (ii) Multiplexing two single-resolution multimedia services allowed for the transmission of more hyper-local services within the available radio resources. (iii) Macro-diversity combining minimized service outage for users requesting local services. (iv)The coverage area for hyper-local services was limited due to Inter-Cell Interference (ICI), and techniques to handle the same could be implemented in the future. (v) Inter-Layer Interference (ILI) poses a persistent challenge in the proposed LHS transmission scheme, but it can be effectively addressed using existing techniques from the literature. Future scope also includes the evaluation of additional performance metrics for our proposed scheme, such as (a) Mean Throughput, which measures the average data rate experienced by users to provide insights into the ‘satisfaction’ level of the multicast users, and (b) Spectral Efficiency to assess how effectively our proposed scheme utilizes system resources.

\bibliographystyle{IEEEtran}
\bibliography{refList}

\vfill

\end{document}